\documentclass[%
 reprint,
 amsmath,amssymb,
 aps,
prl,
]{revtex4-2}

\usepackage{amssymb}
\usepackage{color}
\usepackage{graphicx}
\usepackage{dcolumn}
\usepackage{bm}
\usepackage[toc,page]{appendix}
\usepackage{tikz}
\usepackage{amsmath}
\usepackage[english]{babel}
\usepackage[pdftex,plainpages=false,colorlinks=true,linkcolor=blue, citecolor=blue, urlcolor=blue]{hyperref}

\newcommand{\dx}{{d_{x^2-y^2}}}
\newcommand{\dz}{{d_{z^2}}}

\newcommand{\lno}{{\mathrm{La_3Ni_2O_7}}}

\begin{document}

\preprint{APS/123-QED}


\title{Intertwined Electron Pairing in the Bilayer Two-orbital Kanamori-Hubbard Model: a Unified Picture of Two Superconductivities in $\mathrm{La_3Ni_2O_7}$
}

\author{Shi-Cong Mo}
\affiliation{ Guangdong Provincial Key Laboratory of Magnetoelectric Physics and Devices,  School of Physics, Sun Yat-sen University, Guangzhou, Guangdong 510275, China
}

\author{Yao-Yuan Zheng}
\affiliation{ Guangdong Provincial Key Laboratory of Magnetoelectric Physics and Devices,  School of Physics, Sun Yat-sen University, Guangzhou, Guangdong 510275, China
}

\author{W\'ei W\'u}
\email[Corresponding author: ]{wuwei69@mail.sysu.edu.cn}
\affiliation{ Guangdong Provincial Key Laboratory of Magnetoelectric Physics and Devices,  School of Physics, Sun Yat-sen University, Guangzhou, Guangdong 510275, China
}

\date{\today}


\begin{abstract}
The mechanism of superconductivity in $\mathrm{La_3Ni_2O_7}$ bulk and film superconductors remains actively debated.
Here, we investigate the bilayer two-orbital  Kanamori-Hubbard  model for  $\mathrm{La_3Ni_2O_7}$  using cellular dynamical mean-field theory. 
 We discover two intertwined $s_{\pm}-$ wave superconductivities with distinct physical origins. We show that when the $d_{z^2}$ orbital is under-doped, electron pairing associated to Hund's coupling $J_H$  prevails. As $d_{z^2}$  hole-doping $\delta_z$ increases, a second superconductivity, which is largely insensitive to $J_H$ but exhibiting a critical reliance on the  $d_{z^2}$ - $d_{x^2-y^2}$ hybridization $V$, arises.
 These two primary pairing states exhibit comparable maximum transition temperatures $T_c$ ,  and   evolve from one to the other following a smooth  $T_c$ versus $\delta_z$ relation.
A stark particle-hole asymmetry is observed in the superconducting phase diagram, indicating the crucial role played by the $\gamma-$ band of $d_{z^2}$ orbital in pairing.
Our results present a  picture unifying the two possible pairing mechanisms in  $\mathrm{La_3Ni_2O_7}$ superconductors. We discuss the  implications of our findings to recent experiments.
\end{abstract}


\maketitle

\textit{Introduction-}
The recent discovery of pressure induced superconductivity (SC) in bilayer nickelate $\mathrm{La_3Ni_2O_7}$~\cite{sun2023} represents a significant breakthrough in the field of unconventional superconductivity~\cite{bednorz1986,stewart2011_RMP,maier2011pair,keimer2015quantum,li2019_ni112}. As a new class of nickelate superconductors~\cite{li2019_ni112,mjiang_112,hfchen_112,karp20_112,lu2021magnetic} with enhanced transition temperatures ($T_c$),  $\lno$  has stimulated a surge of experimental
~\cite{jgchen24_nature,yuanhq_24,hhwen_3layer24,jzhao24_4310,dlfeng24,meng2024density,lshu24_usr,xhchen25,li_100gpa,jjzhang_25,zychen2025_odef} and theoretical ~\cite{zhluo23_prl,christiansson2023correlated,qhwang23_frg,yyang_2compon23,fyang23_rpa,cjwu24_hund,zhluo24_qm,zheng2025s,Kuroki24,oh2023type,
lechermann23,wu24_charge,ku24,zhang24_dft,geisler24,ryee2024quenched,jkun24,xtao24,jphu25,  lqiao25_scpma,zylu24_hund,gsu24_tn,liao2023electron,wang2025fermi,ryee2025optimal} studies. While a  large body of theoretical work~\cite{qhwang23_frg,yyang_2compon23,fyang23_rpa,cjwu24_hund,zhluo24_qm,zheng2025s} and recent experimental probes ~\cite{wen_gap,shen2025nodeles}  point to a primary nodeless $S$-wave pairing symmetry, the underlying physical origin of SC  remains a subject of debate. One central, unresolved problem in $\lno$ SC involves resolving the respective contributions of  $\mathrm{Ni}-3d$ $e_g$ doublets (\textit{i.e.} $\dx$ and $\dz$  orbitals) to electron pairing ~\cite{zhluo23_prl,yyang_2compon23,cjwu24_hund,ku24,jkun24,xtao24}.

\begin{figure}[t!]
\includegraphics[scale=0.3]{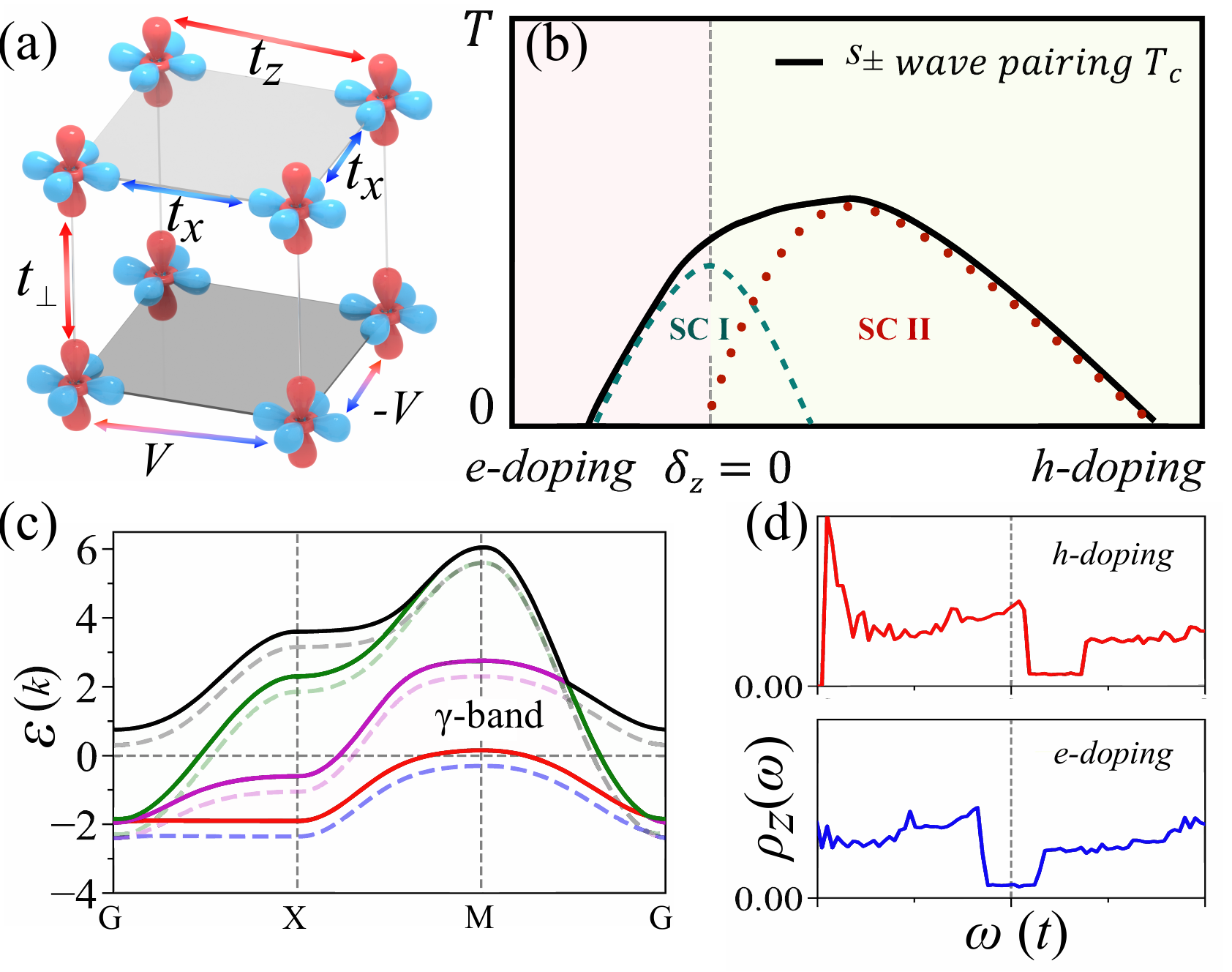}
\caption{Illustration of the bilayer two-orbital model for $\lno$ and its intertwined two superconductivities. \textbf{(a)} Some of the key hopping terms (Sec.A). \textbf{(b)} Schematic of superconducting $T_c$ as a function of $\delta_{z}$. Two superconducting states, SC I and SC II, dominate different doping regimes (\textit{cf.} Fig.~\ref{fig:sc} and main text). Note that competing orders such as the spin density wave or charge density wave are excluded here. $\delta_z = 1- n_z$ denotes the $\dz$ doping level. e-doping: $\delta_z < 0 $ (electron doping), h-doping: $\delta_z > 0 $ (hole doping) \textbf{(c)} Non-interacting Band structure of the 2D BLTO model. Solid  (dashed)  lines show case with (without) SBM. \textbf{(d)} Local density of states (DOS) of the $\dz$ orbital, $\rho_z(\omega)$. Upper (lower) panel  corresponds to case with (without) SBM as shown in (c).
}
\label{fig:illu}
\end{figure}

Unlike cuprates,   $\lno$ has a   $\mathrm{NiO_2}$ bilayer  structure and  both $\dx$ and $\dz$  orbitals are close to Fermi level~\cite{zhluo23_prl}.
Previous studies~\cite{sun2023,zhluo23_prl}  suggest that the appearance of the $\gamma-$ band of $\dz$ orbital at Fermi level, or namely the so-called ``$\sigma-$ bonding band metallization" (SBM)\cite{gmzhang23_cpl,cpl_review}, which causes a boost of the 
 low-energy $\dz$ density of states (DOS)  (see Fig.~\ref{fig:illu}), plays a vital role in the onset of SC.
One of the microscopic theories falling in this category is the ``two-component" theory of superconductivity ~\cite{yyang_2compon23,berg08}, which considers 
 $\dz$ orbitals as the origin where electrons pair,  while $\dx$ electrons, via the $\dx-$ $\dz$ hybridization $V$,  making phase coherence of electrons pairs possible. Nevertheless, there are also studies~\cite{cjwu24_hund,shao2025pairing} suggesting that SBM can be absent in $s_\pm -$ wave pairing in $\mathrm{La_3Ni_2O_7}$. In this scenario, Hund's coupling $J_H$  between the two $e_g$ orbitals is deemed important, which transfers magnetic correlations between inter-layer $\dz$ electrons to inter-layer $\dx$ electrons, driving an SC instability rooted in  $\dx$ orbitals. Obviously, these two scenarios have just opposite views on the role of SBM in SC: The ``two-component" theory requires low-energy  $\dz$ electrons as one of the components of SC, hence SBM is essential for enhancing SC. The latter theory though requires magnetic correlations in $\dz$ orbital to drive $\dx$ electron pairing, but magnetic correlations can come with escalated energy scales. Thus SBM is not a prerequisite. In fact, since SBM delocalizes $\dz$ local moments, it in principle  impedes SC in this scenario.

The debate has been further complicated by recent reports of ambient-pressure SC in $\mathrm{La_3Ni_2O_7}$ films~\cite{hhuang25_film,zxshen25_arpes,zychen25_film,jfhe_arpes,yfnie_film25,li2025enhanced}.
While some experiments found SC~\cite{hhuang25_film,zxshen25_arpes} ($T_c \sim 40$ K) in $\mathrm{(La,Pr)_3Ni_2O_7}$ films in the \textit{absence of a $\gamma$ pocket} on Fermi surface~\cite{zxshen25_arpes}, other groups revealed SC~\cite{zychen25_film,jfhe_arpes} in similar systems claiming the \textit{presence of $\gamma$ -band} at Fermi level~\cite{jfhe_arpes}. A systematic strontium ($\mathrm{Sr}$)  doping experiment on  $\mathrm{La_{3-x}Sr_{x}Ni_2O_7}$ films~\cite{yfnie_film25} found similar $T_c$
 for zero doping ($x=0$) and a relative large hole doping ($x=0.21$), suggesting SC can be enhanced both with and without SBM. Moreover, studies on bulk $\mathrm{La_3Ni_2O_7}$ find SC at very high pressures ($>80 \mathrm{GPa}$)~\cite{li_100gpa}, a regime where SBM is very likely present~\cite{zhang24_dft}. These seemingly contradictory observations, regarding  the role of $\gamma-$ band, challenge current theoretical interpretations.
Resolving this dichotomy and unraveling the relationship between SBM and SC  is therefore  crucial for the understanding of $\mathrm{La_3Ni_2O_7}$ superconductivity.

In this paper, we address this issue by studying superconducting states in a doped bilayer two-orbital (BLTO) Kanamori-Hubbard model for $\lno$ on two dimensional (2D) square lattice~\cite{zhluo23_prl}, using cellular dynamical mean-field theory (CDMFT)~\cite{georges96,kotliar01} (see Sec.A in Supplementary Materials). We identify two distinct $s_{\pm}-$ wave pairing states reigning different parameter regimes. When $\dz$ orbital is electron doped or slightly hole-doped (without SBM), 
 superconducting state (SC I) associated to enhanced $J_H$ can emerge. As $\dz$ orbital is considerably hole doped ($\delta_z > 0 $, with SBM), SC I state is suppressed and a second superconductivity (SC II), which is induced by a moderate hybridization $V$, becomes dominant. In between the two regimes, SC I and SC II pairing coexist and collectively determine superconducting $T_c$, which evolves smoothly with  $\delta_z $ despite pairing type can change. These results
 suggest SC can occur both \textit{in the absence and presence} of $\gamma-$ band at Fermi level in $\lno$ systems, as illustrated in Fig.~\ref{fig:illu}b.
 A stark particle-hole asymmetry regarding $\dz$ orbital  is found in the SC phase diagram, revealing the crucial role of SBM played in superconductivity. 
 Our SC phase diagram  aligns with recent experimental findings on $\mathrm{Sr-}$ doped $\mathrm{La_{3-x}Sr_xNi_2O_7}$ film~\cite{yfnie_film25}.

\begin{figure}[b!]
\includegraphics[scale=0.08]{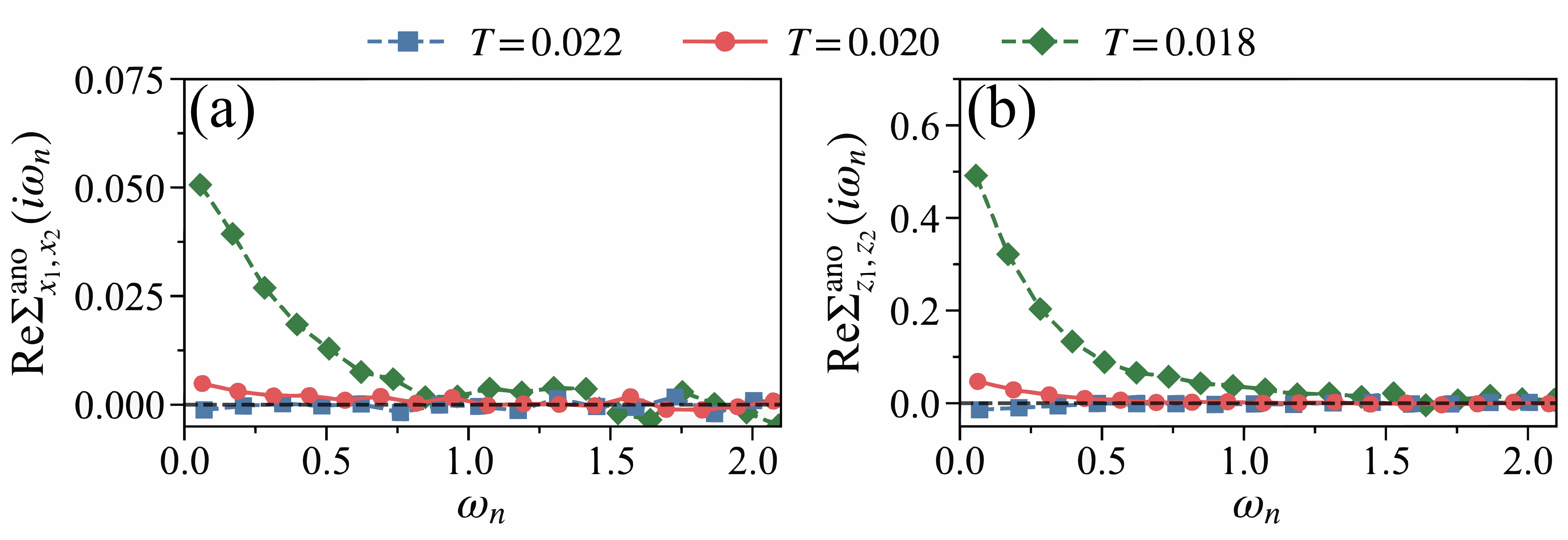}
\caption{Real part of anomalous self-energy  as a function of Matsubara frequency at different temperature $T$.
\textbf{(a)} Inter-layer-intra-orbital component between two $\dx$ orbitals.
\textbf{(b)} The same component between two $\dz$ orbitals.
 Here $V=0.5t, t_{\perp} = -0.8t, U=7t, J_H = 0.2U, n_x=0.6, n_z=0.92$.
}
\label{fig:od}
\end{figure}

\textit{Superconducting phase diagrams - } To tackle the superconducting states, our CDMFT calculations are carried out in Nambu notation where SC order parameter is computed to identify $T_c$. Other long-range orders, such as the  spin-density-wave or charge-density-wave are not considered in our calculation\cite{dlfeng24,meng2024density,lshu24_usr,xhchen25}. As an example, in Fig.~\ref{fig:od} the inter-layer-intra-orbital components of the local anomalous self-energy in Matsubara space for $\dx$ and $\dz$ orbitals, $\mathrm{Re}\Sigma^{ano}_{x1,x2}(i\omega_n)$, and $\mathrm{Re}\Sigma^{ano}_{z1,z2}(i\omega_n)$ are shown at three different temperatures $T$, where one can identify a superconducting $T_c \approx 0.02t$ at given parameters. Note that here since the hybridization $V$ between the two $e_g$ orbitals is finite,  $\mathrm{Re}\Sigma^{ano}_{x1,x2}(i\omega_n)$, and $\mathrm{Re}\Sigma^{ano}_{z1,z2}(i\omega_n)$ would become non-zero simultaneously as SC develops. In our study, electron
pairing occurs mostly between local inter-layer electrons, which generally corresponds to an $s_{\pm}-$ wave superconductivity, like found in previous studies~\cite{qhwang23_frg,zhluo24_qm,zheng2025s}. Repeating this calculation at different parameters we obtain SC phase diagrams, as to be discussed below.

\begin{figure}[t!]
\includegraphics[scale=0.32]{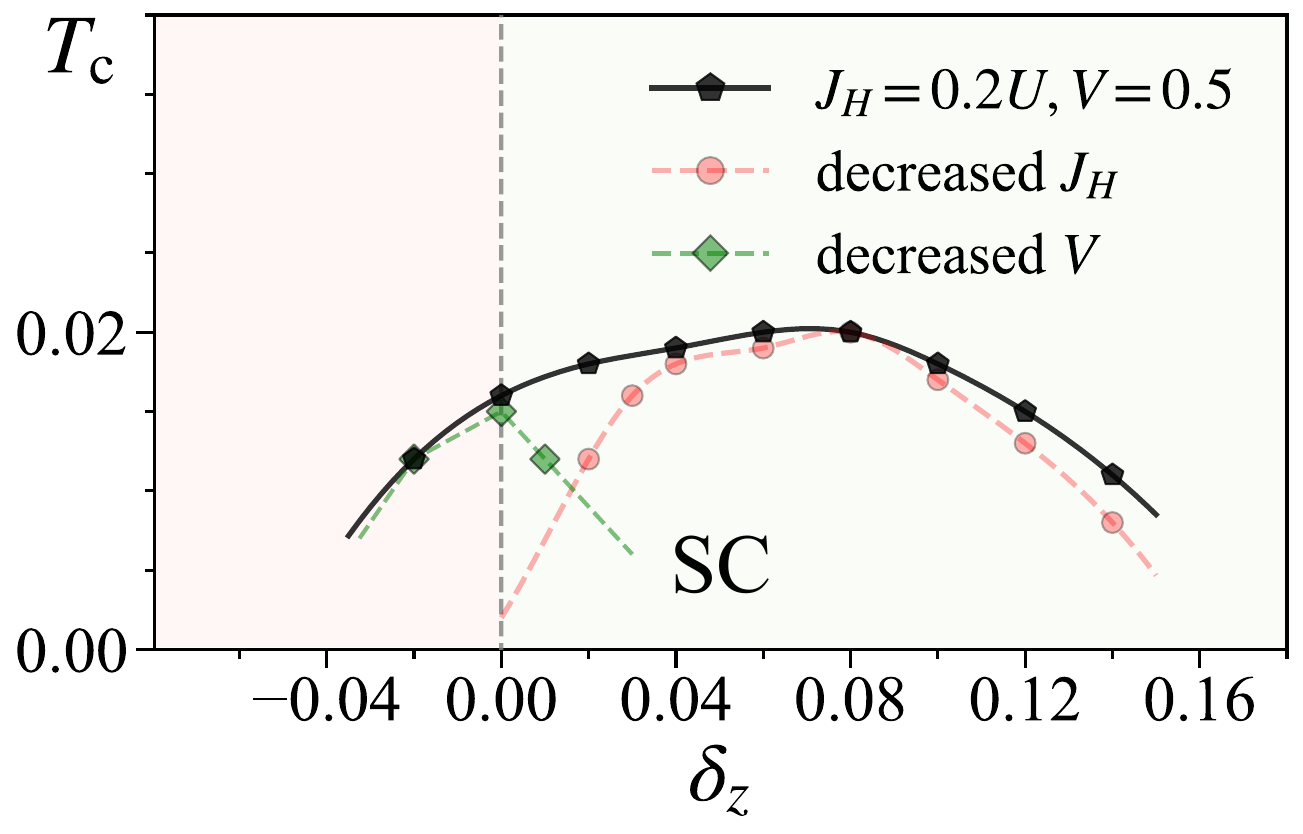}
\caption{Superconducting $T_c$ as a function of $\dz$ doping $\delta_z$.  Solid line displays $T_c$ versus $\delta_z$ at $J_H = 0.2U, V=0.5t$.  For comparison, data for smaller $J_H$ case, $J_H = 0.15U, V=0.5t$ (Dots) and for smaller $V$ case, $V = 0.2t, J_H=0.2U$ (Diamonds) are 
shown by dashed lines. Other parameters are $n_x=0.6, t_{\perp} = -0.8t, U=7t$. At $\delta_z = 0$, the lowest temperature we have accessed is $T= 0.008t$ for $J_H = 0.15U$ (Dots)  where no signature of SC is found (see Sec.E in Supplementary Materials). The lines are guides for the eye.
}
\label{fig:sc}
\end{figure}

 We now investigate the dependence of SC on  $\dz$ hole-doping level $\delta_z$, while keeping all other physical parameters fixed. In this process, the location of $\dz$ bonding band relative to Fermi level will be varied. 
 Fig.~\ref{fig:sc} displays $T_c$ as a function of $\delta_z$ for $J_H = 0.2U$ (Pentagons), where SC is projected to exist in a broad doping range ($-0.04 \lesssim \delta_z  \lesssim 0.18$). We first focus on the hole-doping side ($\delta_z \geq 0$). At half-filling ($\delta_z = 0$), SC occurs with a finite $T_c \approx 0.016t$. As $\delta_z $ increases, a mild enhancement of superconducting 
$T_c$ can be seen, until $\delta_z = 0.08 $ where a maximum $T_c \approx 0.02t$ is obtained. Further increasing  $\delta_z $ reduces $T_c$, and SC can eventually be quenched when $\delta_z \gtrsim 0.18 $. We note that this phase diagram looks different from the dome-shaped hole-doping phase diagram of cuprates~\cite{wu25interplay}, where $T_c \rightarrow 0 $ for hole-doping $\delta \rightarrow 0$. How to understand the evolution of $T_c$ here?  
For large hole dopings ($\delta_z \gtrsim 0.08$), one may attribute the decreasing of $T_c$  to the fading out of correlation effects  upon doping. For small dopings ($0 \leq \delta_z  \lesssim 0.08$), however, the mechanism underlying a slowly changing $T_c$ as a function $\delta_z$  is much less obvious.

To gain insight, Fig.~\ref{fig:sc}
 plots two  additional $T_c$ curves (dashed lines) with modified $V$ and  $J_H$  respectively. 
For the decreased $V$ case (Squares), SC is seen suppressed  as $\delta_z $ increased from zero, whereas for decreased $J_H$ case (Dots), $T_c$ is enhanced with  $\delta_z $ for $\delta_z \lesssim 0.08$. Namely,  hole-doping can play either a detrimental or beneficial role for pairing, depending on the specific values of $V$ or $J_H$.
This finding strongly implies multiple pairing mechanisms are at play in the phase diagram. We propose that the two dashed $T_c$ lines shown here actually separate different reigning regimes of two distinct superconductivities (\textit{cf.} Fig.~\ref{fig:illu}b), a point we will elaborate on below.

\begin{figure}[t!]
\includegraphics[scale=0.22]{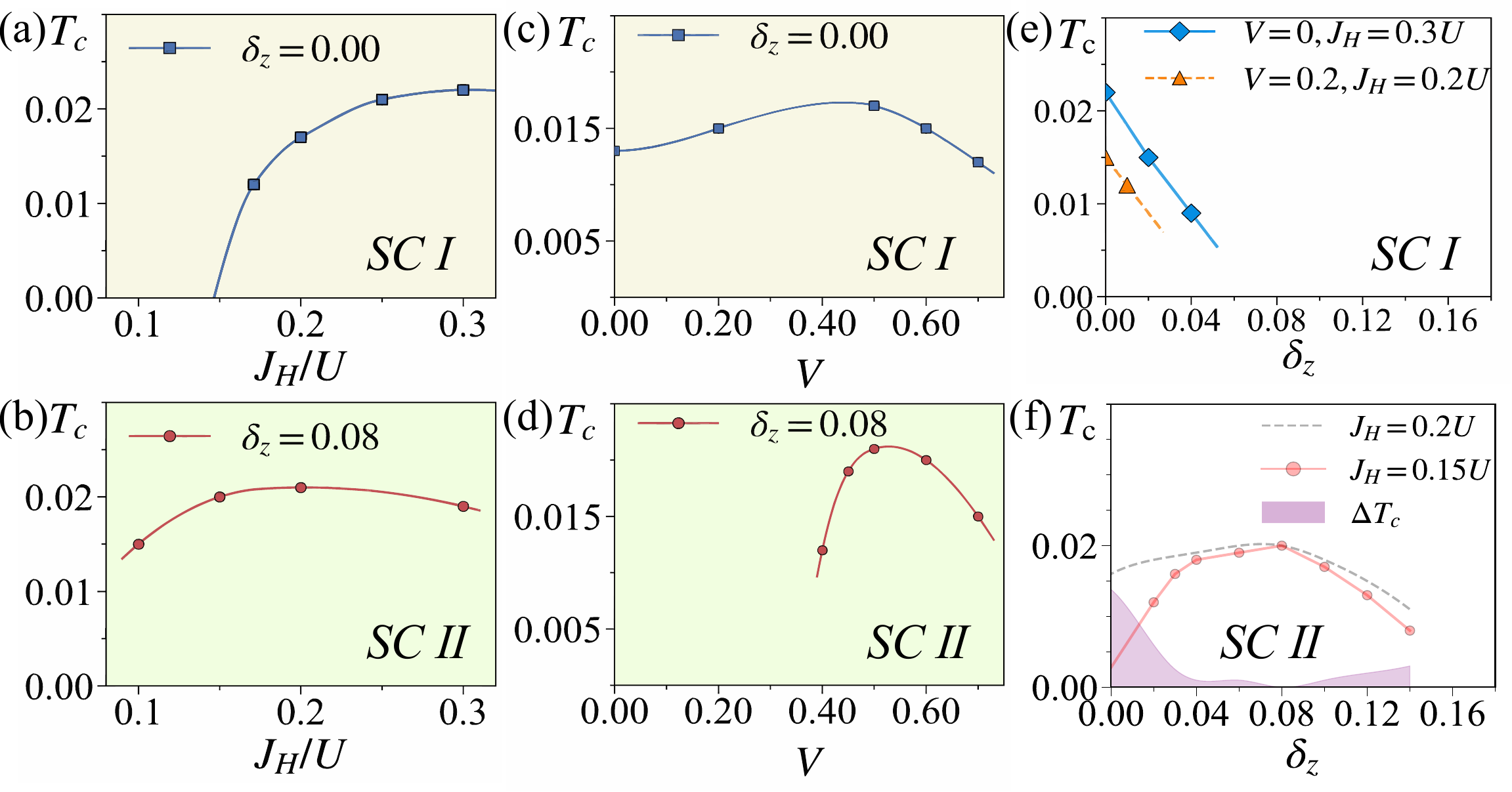}
\caption{Dependence of superconducting $T_c$ on various physical parameters. \textbf{(a)} $T_c$ as a function of Hund's coupling $J_H$ at $\delta_z = 0, V=0.5t$. \textbf{(b)} Same as (a) but at $\delta_z = 0.08$. \textbf{(c)} The $V-$ dependence of $T_c$ at  $\delta_z = 0, J_H = 0.2U$.  \textbf{(d)} 
Same as (b) but at $\delta_z = 0.08$. \textbf{(e)} $T_c$ as a function $\delta_z$ at decreased $V$. Here two $J_H$ cases are shown.
 \textbf{(f)} $T_c$ versus $\delta_z$ at $J_H = 0.15U, V=0.5t$ (Dots). The difference in  $T_c$, $\Delta T_c$ between two cases with $J_H = 0.15U/0.2U$ is  shown in color.
  Other parameters are $n_x=0.6, t_{\perp} = -0.8t, U=7t$. The lines are guides for the eye.  
}
\label{fig:pd}
\end{figure}

\textit{Two superconductivities - }
We now focus on the distinct characteristics of the two SCs. Shown in Fig.~\ref{fig:pd}a and Fig.~\ref{fig:pd}b are $T_c$ as a function of $J_H$ for two different $\delta_z $ cases, $\delta_z =0 $ and $\delta_z = 0.08 $  respectively. For $\delta_z =0 $, superconducting $T_c$ displays a critical growth with $J_H$ around $J_H  \approx 0.15U $,  suggesting electron pairing is intrinsically related to Hund's coupling $J_H$ here. For  $\delta_z = 0.08 $, in contrast, $J_H$ in the same range ($0.1U < J_H < 0.3U$) exhibits a much milder influence on $T_c$.  In fact, at  $\delta_z = 0.08 $, the decrease of $T_c$ with reducing $J_H$ (when $J_H \lesssim 0.2U$ ) is mainly due to the reduced effective mass of $\dz$ orbital, an effect of $J_H$~\cite{medici11} extrinsic to pairing (see Sec.G in Supplementary Materials).
On the other hand, Fig.~\ref{fig:pd}c and Fig.~\ref{fig:pd}d display the $V-$ dependence of $T_c$ for the two representative dopings. At $\delta_z = 0$, $T_c$ exhibits only a slight decrease as $V$ reduced from $V=0.5t$, which is the base value we have used for $V$. Interestingly, here even when $V$ approaches zero $V=0$, a finite $T_c$ can be obtained. This finding indicates that the SC state at $\delta_z = 0$ does not rely on the $\dx-$ $\dz$ orbital hybridization $V$.
 By contrast,  at $\delta_z = 0.08$ ,  a rapid suppression of $T_c$ can be seen as $V$ decreasing from $V=0.5t$, and a critical $V_c \sim 0.35t$  can be inferred where SC vanishes in the zero $T$ limit.

From above study, we learn that with respect to the $J_H$ and $V$ dependence, SC at the two representative $\delta_z$ dopings are fundamentally different. At $\delta_z = 0$, SC exhibits a critical dependence on $J_H$ but weak sensitivity to $V$. The situation is just reversed for the $\delta_z= 0.08$ case, where SC becomes strongly dependent on  $V$ but only weakly on $J_H$.
This observation lead us to conclude that there can be two different superconductivities in the BLTO Kanamori Hubbard model: At small electron- or hole- dopings ($\delta_z= 0$  for example), electron pairing involving Hund's coupling $J_H$ develops primarily~\cite{cjwu24_hund}, which is dubbed as the ``SC I" superconducting state. At large hole-dopings ($\delta_z= 0.08$ for example) electron pairing associated to $\dx-$ $\dz$ orbital hybridization $V$ predominates~\cite{yyang_2compon23}, which is dubbed as the ``SC II" state, as illustrated in Fig.~\ref{fig:illu}b. 

One import issue remaining is that how SC instabilities evolve from  SC I dominated pairing at $\delta_z= 0$, to SC II dominated pairing at large dopings? For intermediate $\delta_z$, there  also  could be coexistence of the two SCs. To disentangle the two pairing mechanisms, we tune  $V$ and $J_H$  to selectively activate/deactivate each SC. For example, when $V$ is decreased from $V=0.5t$ to $V=0.2t$, SC I is preserved (Fig.~\ref{fig:pd}c) while SC II is turned off (Fig.~\ref{fig:pd}d). On the other hand, when $J_H$ is reduced from $J_H = 0.2U$ to $J_H = 0.15U$, SC II is maintained (Fig.~\ref{fig:pd}b) while SC I can be quenched (Fig.~\ref{fig:pd}a).
Along this direction,  in Fig.~\ref{fig:pd}e we fix $V =0.2t$ to study the evolution of SC I upon hole-doping at $J_H = 0.2U$. One can see that as $\delta_z $ becomes finite, $T_c$ drops rapidly from its maximum value of $T_c \sim 0.015t$ at  $\delta_z = 0$ (Triangles), demonstrating the fragility of SC I  against  holes in $\dz$ orbital. Notably, increasing $J_H$ to $J_H = 0.3U$ does not alter this tendency (Diamonds).
Moreover, in Fig.~\ref{fig:pd}f, $T_c$ at $J_H =0.15U$ is plotted to trace the evolution of SC II as a function of $\delta_z$ (Dots) in hole-doped regime. One can see that almost in the same doping range where SC I get destroyed (Fig.~\ref{fig:pd}e),   $T_c$ of SC II increases from vanishingly small ($\delta_z =0 $) to an almost saturating value  $T_c \sim 0.018t$ ($\delta_z =0.04 $). In essence, hole-doping the half-filled $\dz-$ orbital rapidly suppresses SC I  but concurrently enhances SC II, at a nearly equivalent rate.
Considering also the two SCs have comparable maximum $T_c$, this compensatory effect  explains why  $T_c$ evolves rather smoothly with  $\delta_z$ for $ 0 \leq \delta_z \lesssim 0.05$ (Fig.~\ref{fig:sc}), despite underlying pairing mechanism changes type from SC I to SC II. Interestingly, in a recent $\mathrm{Sr}-$ doping experiment, it is also found that $\mathrm{La_{3-x}Sr_xNi_2O_7}$ films exhibit $T_c$ almost unchanged over a range of $\mathrm{Sr}-$ content ($x$), $0 \leq x \lesssim 0.21$~\cite{yfnie_film25}.

\textit{The dependence of  SC on other parameters -} We have identified two SCs, each dominating  a specific range of $\delta_z$.  Determining the precise value of $\delta_z$ where the interacting $\gamma-$ band forms a Fermi pocket is, however, non-trivial. As in  underdoped cuprates~\cite{wu2018pseudogap}, relating  strongly correlated physics to Fermi surface (FS) property can lead to  ambiguousness. Because strong correlation effects can renormalize the geometry of non-interacting FS, and smear the low-energy spectral functions,which blurs the definition of FS itself.
Notwithstanding the uncertainties,
 we note that there is stark particle-hole (PH) asymmetry in our SC phase diagram (Fig.~\ref{fig:sc}), which contains an SC II regime (Dots) broad on the hole-doping side ($\delta_z > 0$) but vanishing on the electron-doping side ($\delta_z <0$) (Sec.C in Supplementary Materials).
We argue that this PH asymmetry demonstrates the direct correlation between the enhancement of  SC II and the appearance of low-energy $\gamma-$ band. Indeed, from a band-structure perspective, $\dz$ orbital has larger DOS on the hole-doped side that essentially originates from the $\gamma-$ band (see Fig.~\ref{fig:illu} and Sec.D in Supplementary Materials).

\begin{figure}[t!]
\includegraphics[scale=0.26]{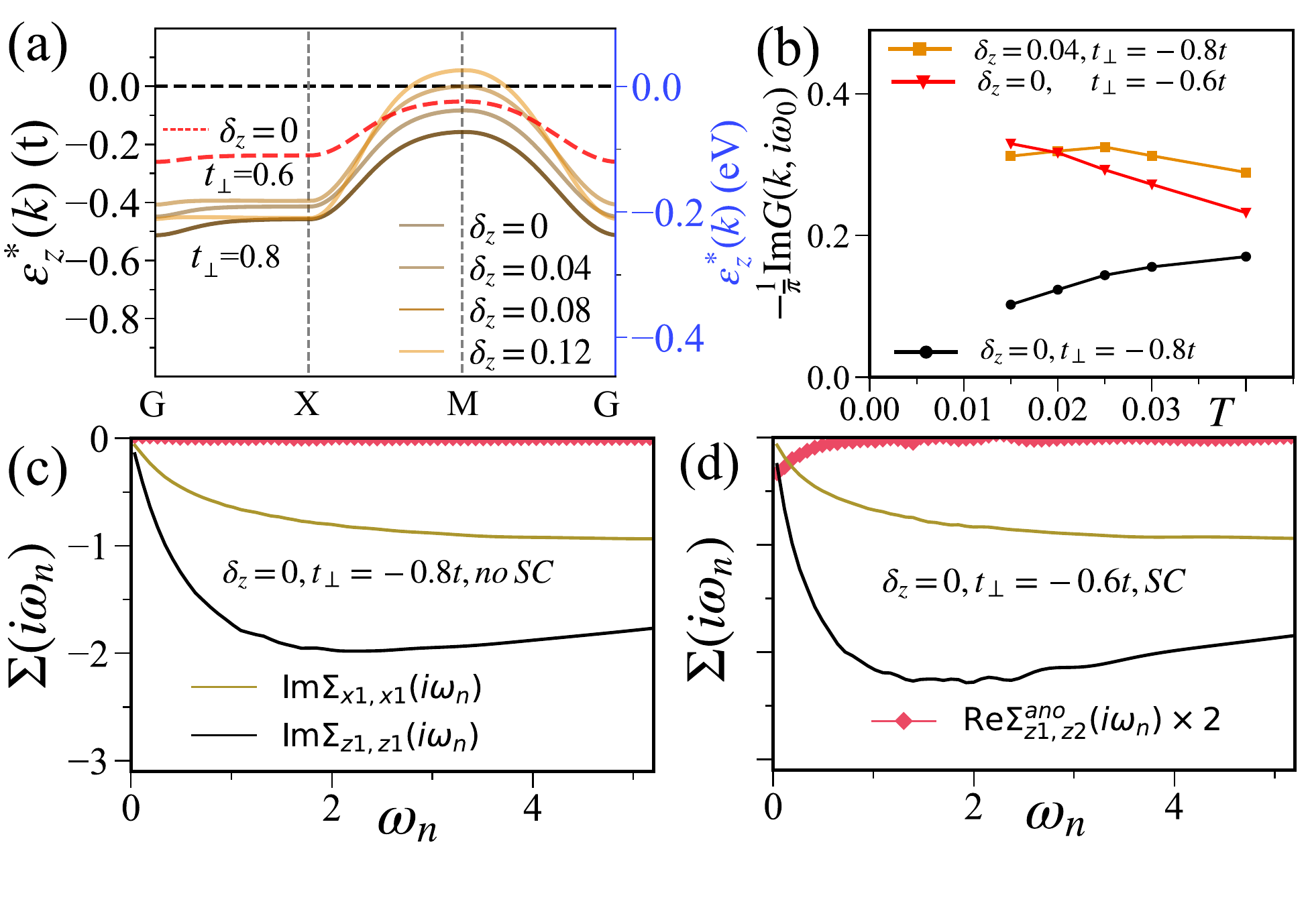}
\caption{$\gamma-$ band and self-energy  in SC II pairing. \textbf{(a)}
 $\gamma-$ band of $\epsilon^{*}_{z}(k)$  along high-symmetry paths in the Brillouin zone for several $\delta_z$ at 
 $t_{\perp} = -0.8t$ (solid lines), and for $\delta_z = 0, t_{\perp} = -0.6t$ (dashed line). Refer to Fig.~\ref{fig:pd}f for the  $T_c$ versus $\delta_z$ curve. Here $T=0.02t$. \textbf{(b)} Approximate spectral function $\tilde{A}(\mathbf{k}, \omega_0) \equiv -\frac{1}{\pi} \mathrm{Im}G(\mathbf{k}, i\omega_0)$  as a function of temperature $T$ at ${\mathbf{k}=(\pi,\pi)}$ for three cases with different $\delta_z$ or $t_{\perp}$. \textbf{(c)} Anomalous inter-layer self-energy $\mathrm{Re}\Sigma^{ano}_{z1,z2}(i\omega_n)$ and normal intra-layer self-energies $\mathrm{Im}\Sigma_{x1,x1}(i\omega_n)$, $\mathrm{Im}\Sigma_{z1,z1}(i\omega_n)$ as functions of $\omega_n$ at $\delta_z = 0, t_{\perp} = -0.8t, T=0.012t$. In this case, no finite anomalous self-energy (Diamonds) is found. \textbf{(d)} Same as (c) but $t_{\perp}$ is changed into $t_{\perp} = -0.6t$. Since $T < T_c $ , here the real part of the anomalous self-energy is finite, it is in SC state. Other parameters are $J_H = 0.15U, V=0.5t$.  }
\label{fig:sigma}
\end{figure}

To further support this point, we notice that the zeros of the modified dispersion relation  $\epsilon^{*}(k)= \epsilon(k) -\mu+ \mathrm{Re} \Sigma(k,\omega=0) $ define the (quasi-) poles of the interacting Green's functions, and by $\epsilon^{*}_{z}(k)  = Z\epsilon^{*}(k) $ the  band renormalization effect can also be accounted. Here $\mathrm{Re} \Sigma(k,\omega=0) $ is the real-part of the self-energy and $1/Z$ is  the renormalization factor of a certain energy band (Sec.A).  Thus $\epsilon^{*}_{z}(k)$ in general traces the center of a band in energy. Fig.~\ref{fig:sigma}a shows that at $J_H = 0.15U , \delta_z =0$ (where  SC II is suppressed),  the $\gamma$-band of $\epsilon^{*}_{z}(k)$ is distant from Fermi level. Increasing $\delta_z$ drives the band toward Fermi level, concomitant with a rise in $T_c$ ($\textit{cf.}$ Fig.~\ref{fig:pd}f). At $\delta_z=0.08$, the $\gamma-$ band center crosses Fermi level where a maximum $T_c \sim 0.02t$  was found. 
As one may expect, beyond $\delta_z$, SC II should also be dependent on other parameters that affect the energy of $\gamma-$ band. We confirm that at small $\delta_z$, decreasing $|t_{\perp}|$ can boost SC II pairing by moving $\gamma-$ band closer to Fermi level. For instance, at 
 $J_H = 0.15U, \delta_z = 0$, a greatly enhanced $T_c \approx 0.015t$ can be reached if $t_{\perp}$ is changed from  $t_{\perp} = -0.8t$ to $t_{\perp} = -0.6t$ (dashed line in Fig.~\ref{fig:sigma}a). The comparison between the anomalous and normal self-energies of the two $t_{\perp}$ cases can be found in Fig.~\ref{fig:sigma}c and Fig.~\ref{fig:sigma}d. Here the $t_{\perp} = -0.6t$ case has a more correlated $\dz$ orbital (a larger $|\mathrm{Im} \Sigma_{z1,z1}(i\omega_n)|$ at small $\omega_n$)  than the $t_{\perp} = -0.8t$ case. Therefore, the  suppression of SC II at $\delta_z = 0,t_{\perp} = -0.8t $ in SC phase diagram (Fig.~\ref{fig:pd}f) is  attributable not to overly strong electron correlations in the half-filled $\dz$ orbitals, but as we argue, rather to the $\gamma-$ band positioned far away from Fermi level.

We notice that for certain parameters, $\delta_z = 0 , t_{\perp} = -0.6t$ in Fig.~\ref{fig:sigma}a (dashed line) for example, a finite $T_c \approx 0.015t$ for SC II is present despite its $\gamma-$ band of $\epsilon^{*}_{z}(k)$ has no zeros. This can be attributed to the interaction effects  smearing $\gamma-$ band, which place spectral weight at Fermi level without a sharp quasi-particle pole on FS, like found in an experiment in $\lno$ films~\cite{jfhe_arpes}. To demonstrate this,  we study the quantity $\tilde{A}(\mathbf{k}, \omega_0) \equiv -\frac{1}{\pi} \mathrm{Im}G(\mathbf{k}, i\omega_0)$ at the $\gamma-$ band top [${\mathbf{k}=(\pi,\pi)}$]. This quantity  approximate the spectral function at Fermi level, $A(\mathbf{k}, \omega=0)$,
 and  becomes exact spectral function $\tilde{A}(\mathbf{k}, \omega_0) = 
A(\mathbf{k}, \omega = 0)$ as $T \rightarrow 0$. Fig.~\ref{fig:sigma}b shows that at small $T$, the case with suppressed SC II displays a strongly reduced  $\gamma-$ band spectral weight at Fermi level (Dots), while cases with enhanced SC II (Squares and Triangles) exhibit a greatly increased $\tilde{A}(\mathbf{k}, \omega_0)$. This result again unequivocally strengthen the premise that SC II pairing is enhanced by $\gamma-$ band spectral weight approaching Fermi level.

\begin{figure}[t!]
\includegraphics[scale=0.35]{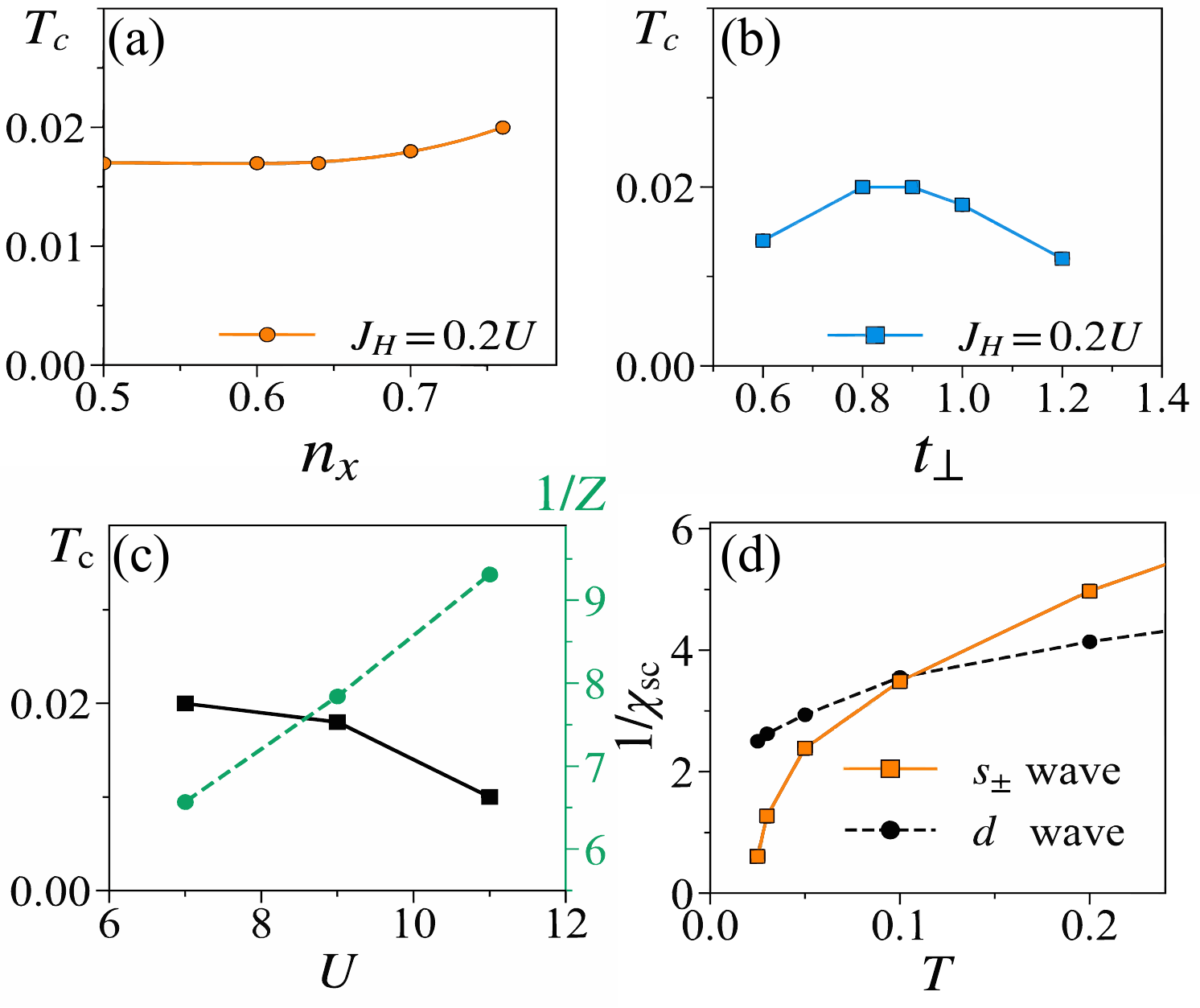}
\caption{Dependence of superconducting $T_c$ on various physical parameters and comparison between $s_{\pm}-$ and $d -$ wave pairing susceptibilities. \textbf{(a)}
 $T_c$ as a function of  $n_x$, the filling factor of $\dx$ orbital, at
$J_H = 0.2U$. Here $\delta_z = 0 (n_z = 1.0) , V=0.5t, t_{\perp} = -0.8t$ \textbf{(b)} The $t_{\perp}$ dependence of $T_c$ at $\delta_z = 0.08 , J_H = 0.2U, V=0.5t, n_x = 0.6$.
\textbf{(c)} $T_c$ and $1/Z_{\gamma}$ as a function of Hubbard $U$ at fixed $J_H = 1.4t$. Other parameters are $\delta_z = 0.08, V = 0.5t, t_{\perp}=-0.8t$. $1/Z_{\gamma}$ is obtained at $T=0.015t$ in the normal state.  \textbf{(d)}  Inverse of the pairing susceptibility $1/\chi_{sc}$ as a function of temperature $T$. Here $ n_x = 0.6, \delta_z = 0.08, J_H = 0, V=0.5t,t_{\perp}=-0.8t$ is used. All results are obtained by $1\times 4-$ orbital CDMFT, except in (d) where $2\times 2 \times 4-$ orbital CDMFT is used.}  
\label{fig:vsu}
\end{figure}

\textit{Robustness of SC I and SC II pairing -} Besides  $\delta_z$, $V$, and $J_H$,  we also  investigate the impacts of other key parameters on SC to ascertain the robustness of the two SCs. Fig.~\ref{fig:vsu}a displays $T_c$  as a function of $n_x$ at ($J_H = 0.2U, \delta_z = 0$), showing weak dependence of $T_c$ on $n_x$. This validates the generality of our study on SC I pairing at $n_x = 0.6$. For SC II at large $\delta_z$,  the dependence of $T_c$ on $\dz$ inter-layer hopping $t_{\perp}$ is rather quantitative, as revealed in  Fig.~\ref{fig:vsu}b. Note that
besides shifting the $\gamma-$ band, $t_{\perp}$ also determines the inter-layer magnetic exchange coupling $J_{\perp} \sim t_{\perp}^2/U$~\cite{wu24_charge}.
In experiments, the strongly correlated nature of $\gamma-$ band in $\lno$ is revealed, whose mass renormalization factor can be as large as $(6 \sim 8)$ ~\cite{yang_orbital24}.
Here, we fix $J_H = 1.4t $ and vary $U$ to change the mass renormalization factor  $1/Z_{\gamma}$ of $\gamma-$ band  of $\dz$ orbital (Sec.A). 
Fig.~\ref{fig:vsu}c displays that SC II pairing at $\delta_z = 0.08$ can persist with an extremely correlated $\dz$ orbital ($1/Z_{\gamma} \sim 9.3 $ at $U = 11$).
Finally, we investigate  possible $d-$ wave pairing  in the system  by $2\times2 \times 4 -$ orbital CDMFT calculations using Hirsch-Fye quantum Monte Carlo (HFQMC)~\cite{hirsch1986monte} impurity solver (Sec.A). Owing to the sign-problem of HFQMC, here $J_H=0$ is used. In Fig.~\ref{fig:vsu}d, the inverse of the pairing susceptibility $1/\chi_{SC}$ for $d-$ wave and $s_{\pm}-$ wave pairing is plotted at $\delta_z = 0.08$, where an $s_{\pm} -$ wave pairing instability is found ($1/\chi_{SC} \rightarrow 0$), in agreement with above CDMFT+CTQMC result. The $d-$ wave pairing, in contrast, seems absent in our study (of which $1/\chi_{SC}$ does not extrapolate to zero at finite $T$). We have also checked $\delta_z = 0$ case, where $d-$ wave SC is also not found .

\textit{Discussion -}
Our results can reconcile the apparent paradox that SC emerges in $\lno$ superconductors both with~\cite{zychen25_film,jfhe_arpes,li_100gpa} and without~\cite{hhuang25_film,zxshen25_arpes} a $\gamma-$ pocket, as they are now considered representing two distinct dominant pairing states (SC I and SC II).
As revealed by DFT calculations~\cite{geisler24}, pristine $\lno$ films on SLAO substrate host a $\gamma-$ band below Fermi level. The $\mathrm{Sr}-$ doping, nevertheless, introduces holes and can elevate the $\gamma-$ band to form a pocket on Fermi surface~\cite{shi2025effect,hu2025electronic,shi2025theoretical,yue2025correlated}. The possible variance in $\mathrm{Sr}-$ doping levels across film experiments may explain the dichotomy on the presence/absence of $\gamma-$ pocket, which could eventually lead to the occurrence of two distinct SCs. Notably, a recent systematic $\mathrm{Sr-}$ doping study on $\mathrm{La_{3-x}Sr_xNi_2O_7}$ films~\cite{yfnie_film25} reports a weak dependence of $T_c$ on $\mathrm{Sr-}$ contend $x$ for small $x$, yielding a phase diagram remarkably similar to ours (solid line in Fig.~\ref{fig:sc}). 
 A key question then is how the $\mathrm{Sr-}$ doping level can be related to $\delta_z$ and the low-energy $\gamma-$ band spectral weight? Holes doped into the system occupy both $\dx$ and $\dz$ orbitals~\cite{wu24_charge,dong2024visualization}. Our calculation indicates that holes went to $\dx $ orbital only quantitatively affects SC, whereas those in $\dz$ orbital can directly shift  $\gamma-$ band, governing the transition between SC I and SC II. Hence, more accurate measuring of $\gamma-$ band spectral weight at Fermi level, as well as probing the strength of Hund's coupling $J_H$ 
are  crucial for clarifying the pairing nature in future experiments~\cite{wang2025fermi,shao2025pairing}.

Finally, while we consider the SC II pairing identified here largely linked to the ``two-component" SC proposed in Ref.~\cite{yyang_2compon23} , and the SC I state to the Hund scenario proposed in Ref.~\cite{cjwu24_hund}, significant difference in pysical properties, particularly those involving SC coexistence and quantum fluctuations, may present.
Future research is needed to elucidate more natures of the two pairing states and the interplay between them.

\section{Acknowledgment} W.W is indebted to Yi-feng Yang for insightful comments on the manuscript. We are grateful to Fan Yang, Congjun Wu, Zhong-Yi Lu, Guang-Ming Zhang,  Xunwu Hu,  Dao-Xin Yao and Meng Wang for useful discussions. This work is supported by the National Natural Science Foundation of China (Grants  No.12494594, No.12274472). We also thank the support from the Research Center for Magnetoelectric Physics of Guangdong Province (Grants No. 2024B0303390001) and Guangdong Provincial Quantum Science
Strategic Initiative (Grant No. GDZX2401010).

\clearpage
\appendix
\renewcommand{\thesection}{S\arabic{section}}
\renewcommand{\thefigure}{S\arabic{figure}} 
\setcounter{figure}{0} 

\begin{center}

    \section{\large Supplementary Materials}

\end{center}

\section*{A\quad Hamiltonian and CDMFT Methods}

We study the BLTO  Kanamori-Hubbard  model~\cite{zhluo23_prl}  on two-dimensional square lattice defined as $H = H_0 +H_U$,

\begin{eqnarray}
H_0 &=\sum_{\alpha l , \beta m}^{ij \sigma} -t_{\alpha l, \beta m}^{ij} c_{i l \alpha  \sigma}^{\dagger}c_{j  m \beta \sigma}   \nonumber
- \sum_{ \alpha,l}^{i\sigma} (\epsilon_{\alpha }+ \mu) n_{ i l \alpha \sigma} \nonumber  \\  
H_{U} &= \sum_{(\alpha\sigma)< (\beta\sigma')}^{il} U_{\alpha\beta}^{\sigma\sigma'} n_{il\alpha\sigma}n_{il\beta\sigma'}\nonumber \\
&-\sum_{\alpha\neq\beta}^{il}Jc_{il\alpha\uparrow}^{\dagger}c_{il\alpha\downarrow}c_{il\beta\downarrow}^{\dagger}c_{il\beta\uparrow} \\
&+\sum_{\alpha\neq\beta}^{il}Jc_{il\alpha\uparrow}^{\dagger}c_{il\alpha\downarrow}^{\dagger}c_{il\beta\downarrow}c_{il\beta\uparrow} \nonumber
\end{eqnarray}

where $t_{\alpha l,\beta m}^{ij}$ enumerates hopping amplitudes
between layer $l$, site $i$, orbital $\alpha$, and layer $m$, site
$j$, orbital $\beta$. $\sigma/\sigma^{\prime}$ indicate spins.
For brevity,  $x,z$ are used to label $d_{x^{2}-y^{2}}$ and $d_{z^{2}}$ orbitals respectively.
Here, we fix the nearest neighbor (NN) intra-layer hoppings as  $t_{xl,xl}^{\left\langle i,j\right\rangle }\equiv t_{x}=t\equiv1$, $t_{zl,zl}^{\left\langle i,j\right\rangle } \equiv t_{z}=0.25t$, and next-nearest neighbor (NNN) intra-layer hoppings as $t_{x}^{\prime} = -0.15t $, $t_{z}^{\prime} = 0.05t $~\cite{zhluo23_prl, zheng2025s}. Other important hopping terms include the inter-layer hopping between $\dz$ orbitals,  $t_{z1,z2}^{i,i}\equiv t_{\perp}$, and intra-layer-inter-orbital hybridization $t_{xl,zl}^{\left\langle i,j\right\rangle }\equiv \pm V $ (see Fig.~\ref{fig:illu}a). $\epsilon_{\alpha}$ denotes the orbital energies and $\mu$ is the chemical potential.
In Kanamori-Hubbard Hamiltonian
$U_{\alpha=\beta}^{\sigma\neq\sigma^{\prime}} = U $ , 
$U_{\alpha\neq\beta}^{\sigma\neq\sigma^{\prime}} = U^{\prime} = U-2J $, and $U_{\alpha\neq\beta}^{\sigma \sigma} = U^{\prime} - J$ are the intra-orbital
and inter-orbital Hubbard interactions, $J$
is the Hund's coupling. To capture the essential physics of $\lno$ systems, $V=0.5t$, $t_{\perp} = -0.8t$, $U = 7t$, $J = 0.1 \sim 0.3 U$ are typically used. For the average densities,  $n_x \equiv (1/2N) \sum_{il\sigma} \langle n_{x,il\sigma} \rangle  \sim 0.5$, $n_z \equiv (1/2N)\sum_{il\sigma} \langle n_{z,il\sigma} \rangle \sim 1.0$ are  used, $N$ is the number of unit cells, each containing two $\dx$ and two $\dz$ orbitals. For theoretical exploration, these parameters are all varied in our study to examine their influence on SC. As demonstrated in the main text,  our findings do not depend on the specific choice of these physical parameters.

\begin{figure}
  \begin{center}
    \includegraphics[width=0.5\columnwidth]{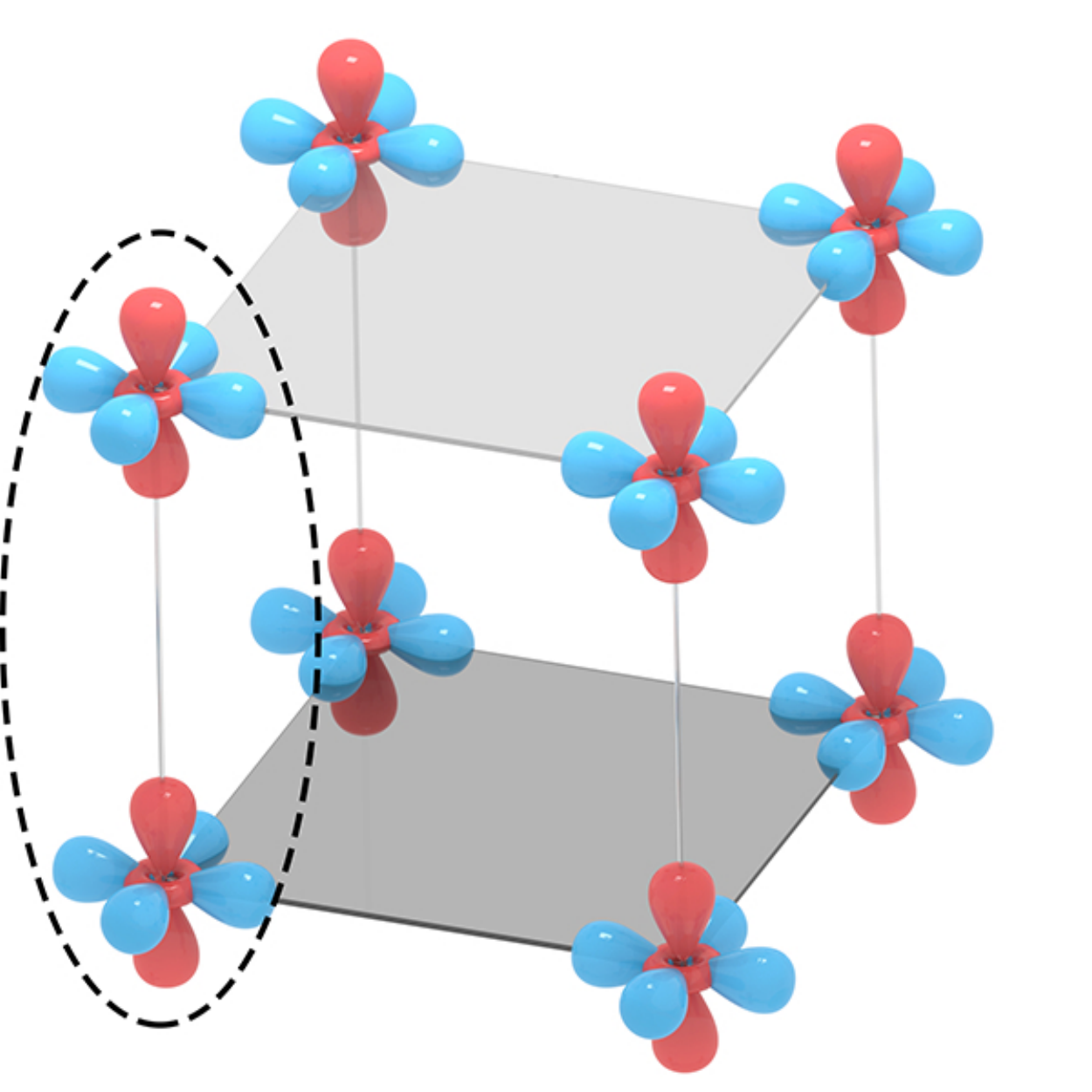}
  \end{center}
  \caption{
    Geometries of the CDMFT clusters used in this work. 
    }
\label{fig:geo}    
\end{figure}

Our results for the two-dimensional BLTO Kanamori-Hubbard model are obtained using the cellular
dynamical mean-field theory (CDMFT)~\cite{kotliar01}, which is a cluster extension of the dynamical mean-field theory (DMFT)~\cite{georges96}.
To obtain the main results, we use $1\times 4-$ orbital effective cluster (namely, one vertical unit cell of the bilayer $\mathrm{NiO_2}$ plane containing  two $\dx$ and two $\dz$ orbitals ), while for the $d-$ wave pairing result in Fig.~\ref{fig:vsu}d, a $2\times 2 \times 4-$ orbital cluster is used (Fig.~\ref{fig:geo}).
  The CDMFT method captures short-ranged spatial 
correlations within the cluster, as well as the temporal fluctuations, exactly. The longer range spatial correlations are then
taken into account by a dynamical mean-field, which can be represented by a momentum- and frequency- dependent Weiss 
field $g_0(k, i\omega_n)$. The effective cluster impurity problem defined by $g_0(k, i\omega_n)$ is solved by an open source implementation  of the continuous time quantum Monte Carlo method (CTQMC), TRIQS3.2~\cite{CTHYB2016,TRIQS2015}  (for $1\times 4-$ orbital cluster), or the Hirsch-Fye quantum Monte carlo method~\cite{hirsch1986monte}( for $2\times 2 \times 4-$ orbital cluster).
We have carefully benchmarked our results from CDMFT+CTQMC and CDMFT+HFQMC, comparison with numerical exact determinant quantum Monte Carlo (DQMC) at high temperatures is also performed, where excellent agreements between those results are obtained, see Sec.B.

In our  $1\times 4-$ orbital CDMFT, self-energies are local in space, $\Sigma_{\alpha l, \beta m}(\mathbf{k}, i\omega_n) \equiv \Sigma_{\alpha l, \beta m}( i\omega_n)  $. Thus, $\Sigma_{z1,z1}( i\omega_n)$ denotes local self-energy between two $\dz$ orbitals at layer-1 (intra-layer intra-orbital self-energy), and $\Sigma_{z1,z2}( i\omega_n)$ the local self-energy between two $\dz$ orbitals at different layers (inter-layer intra-orbital self-energy). In our calculations, the  band renormalization factor of $\gamma-$ band is approximated by $1/Z_{\gamma} = 1- {\mathrm{Im}\Sigma_{\gamma}(i\omega_0)}/{\omega_0}$, where $\Sigma_{\gamma}(i\omega_0)$ is the self-energy of the bonding band of $\dz$ orbital that obtained by diagonalizing  $\Sigma_{\alpha l, \beta m}(\mathbf{k}, i\omega_n)$.

\begin{figure}
  \begin{center}
    \includegraphics[width=0.9\columnwidth]{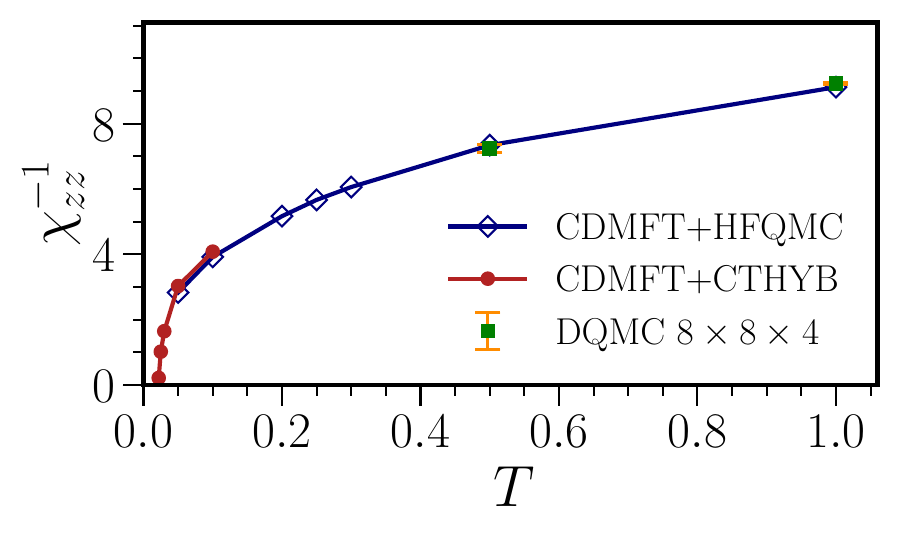}
  \end{center}
  \caption{
    The inverse pairing susceptibility $1/\chi_{zz}$ as a function of temperature $T$. Here $\chi_{zz}$ is the inter-layer-intra-orbital  component of the $s_{\pm}$ pairing  susceptibility  within the $d_{z^2}$ orbitals. Here  $t_x \equiv t = 1, t_x^{\prime} = -0.15t, t_z = 0.25t, t_z^{\prime} = 0.05t, J_H = 0.2U, V=0.5t, t_{\perp} = -0.8t, n_x = 0.6, n_z = 0.92$, see also main text for the definition of the parameters.  $1/\chi_{zz} \rightarrow 0 $  at small $T$ indicates the occurrence of a superconducting instability.
      Due to the restrictions of the sign problem in HFQMC and DQMC, we have neglected the spin-flip and 
     pairing-hopping terms in $J_H$, as to utilize the benchmark. Note that in the main text, we perform calculations considering all terms in $J_H$.
    }
\label{fig:bench}    
\end{figure}

We have used the cellular dynamical mean-field theory (CDMFT)~\cite{kotliar01} to solver the  two-dimensional bilayer two-orbital Kanamori-Hubbard model. The dashed line in the following Fig.~\ref{fig:geo} outlines the $1 \times 4-$ orbital (one unit cell, four orbitals) cluster we have used for obtaining the main results, where an open source implementation (TRIQS3.2, CTHYB) of the hybridization expansion version of the continuous time quantum Monte Carlo method (CTQMC)~\cite{CTHYB2016,TRIQS2015} is used to solver the effective impurity problem. To exam the $d-$ wave pairing, we have used a cluster with four unit cells, like depicted by the  following $2 \times 2 \times 4-$ orbital cluster in Fig.~\ref{fig:geo}. This effective impurity cluster is solved by the Hirsch-Fye quantum Monte Carlo solver~\cite{hirsch1986monte} at Hund's coupling $J_H =0$. Like used in the main text, we fix the intra-layer intra-orbital hoppings through out this Supplementary Materials as $t_x \equiv t = 1, t_x^{\prime} = -0.15t, t_z = 0.25t, t_z^{\prime} = 0.05t$.

\section*{B\quad Benchmarking the CDMFT data}

In Fig.~\ref{fig:bench}, we show results of the inverse pairing susceptibility $\chi_{zz}^{-1}$  using  $1 \times 4-$ orbital cluster. 
Results from CTQMC and HFQMC solvers exhibit excellent agreement in the overlapping regime. At high temperatures, CDMFT+HFQMC result is agreed by the numerical exact determinant quantum Monte Carlo (DQMC) simulation.

\begin{figure}
  \begin{center}
    \includegraphics[width=1\columnwidth]{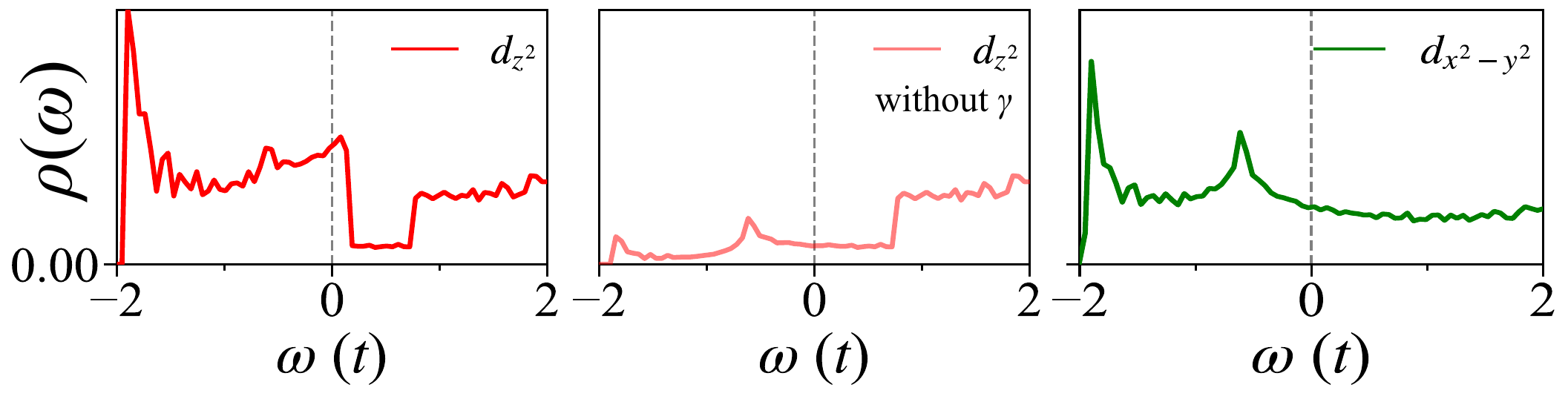}
  \end{center}
  \caption{
    Non-interacting density of states $\rho(\omega)$ of the 2D BLTO model. Here  $  V=0.5t, t_{\perp} = -1.3t$. In the middle panel, we plot DOS with $\gamma-$ band contribution taken out, where the stark PH asymmetry around $\omega=0$ is largely removed.
    }
\label{fig:dos}    
\end{figure}

\section*{C\quad Non-interacting Density of states associated to the $\gamma-$ band}

The bonding band ($\gamma-$ band) of $\dz-$ orbital contributes mainly the density of states (DOS) of $\dz-$ orbital at negative energies, which can gives rise to a strong particle-hole (PH) asymmetry in the low-energy DOS when its band top is in vicinity of the Fermi level, as shown in Fig.~\ref{fig:dos}. In the main text, we have argued that this PH asymmetry essentially leads to a PH asymmetric superconducting phase diagram, as shown in Fig.~\ref{fig:sc} in the main text.

\begin{figure}
  \begin{center}
    \includegraphics[width=1\columnwidth]{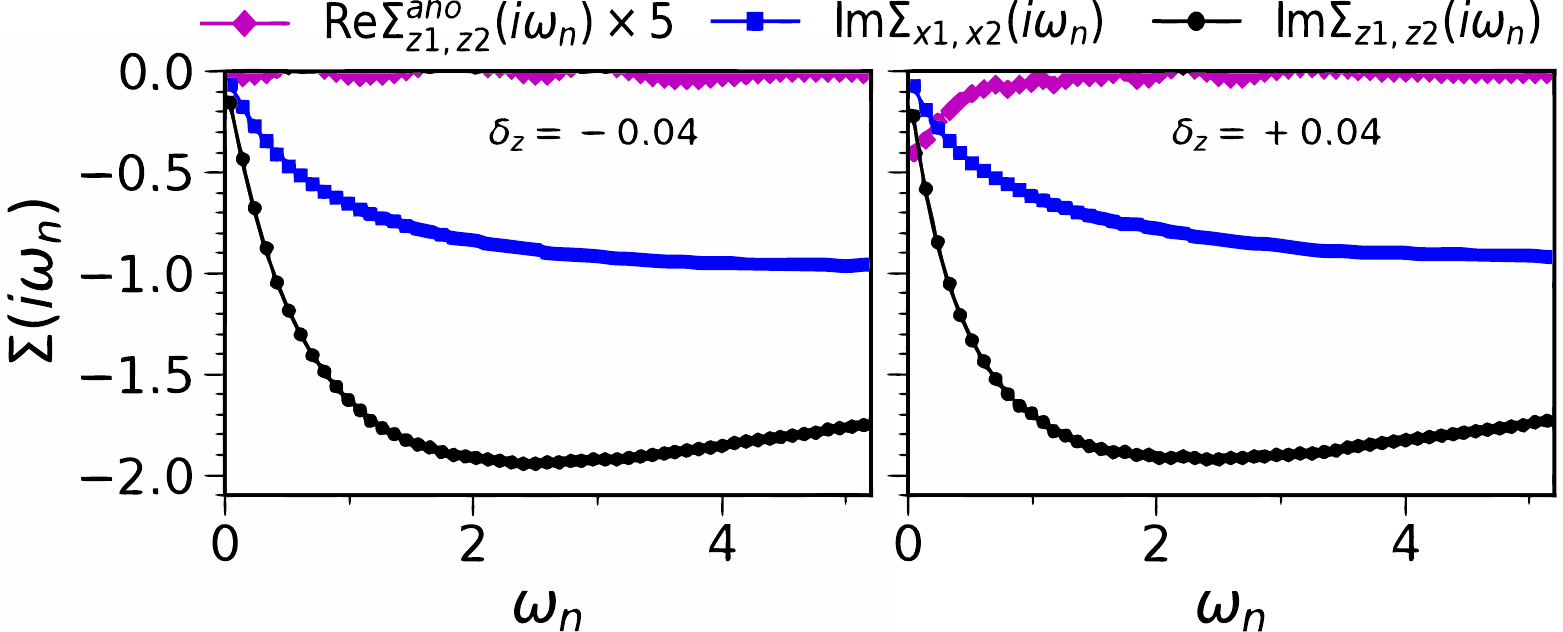}
  \end{center}
  \caption{
    Normal and anomalous self-energies $\Sigma(i\omega_n)$ as  functions of  Matsubara frequency  $\omega_n$ at $4\%$ electron doping and  $4\%$ hole doping. 
    $\mathrm{Re \Sigma_{z1,z2}^{ano}(i \omega_n)}$ is the real part of the local anomalous self-energy between two inter-layer $\dz$ orbitals.
    $\mathrm{Im \Sigma_{x1,x2}(i \omega_n)}$ is the imaginary part of the local normal self-energy  between two inter-layer $\dx$ orbitals.
    $\mathrm{Im \Sigma_{z1,z2}(i \omega_n)}$ is the imaginary part of the local normal self-energy  between two inter-layer $\dz$ orbitals.
    For clarity, the anomalous self-energy is enlarged by five times.
   Here $  V=0.5t, t_{\perp} = -0.8t, U=7t, J_H = 0.15U, T = 0.015t$.
    }
\label{fig:ano}    
\end{figure}

\section*{D\quad $\dz$ particle-hole asymmetry of SC II pairing}

In the main text, we have shown stark particle-hole asymmetry in respect of $\dz-$ orbital doping associated to SC II pairing. Below we present more details about the self-energies at two opposite dopings, $\delta_z = + 0.04$ and  $\delta_z = - 0.04$ at $J_H=0.15U, U=7t$ to elucidate this point. Fig.~\ref{fig:ano} shows that, the $\dz$ inter-layer self-energy  $\mathrm{Im \Sigma_{z1,z2}(i \omega_n)}$ (Dots) of 
$\delta_z = + 0.04$  are larger in magnitude at low-energies. Because hole-doping elevates the $\gamma-$ band closer to Fermi energy,  causing stronger scattering effects at $\delta_z = + 0.04$. For the anomalous self-energy, the electron doping case  $\delta_z = - 0.04$ has vanishing $\mathrm{Re} \Sigma_{z1,z2}^{ano}(i \omega_n)$, suggesting no SC II is found at given $T$. In contrast, for the hole-doping case, finite anomalous self-energy is found since $T = 0.015t$ is lower than its $T_c \approx 0.018t$.

\section*{E\quad SC II pairing without $\sigma-$ bonding band metallization}

\begin{figure}
  \begin{center}
    \includegraphics[width=1\columnwidth]{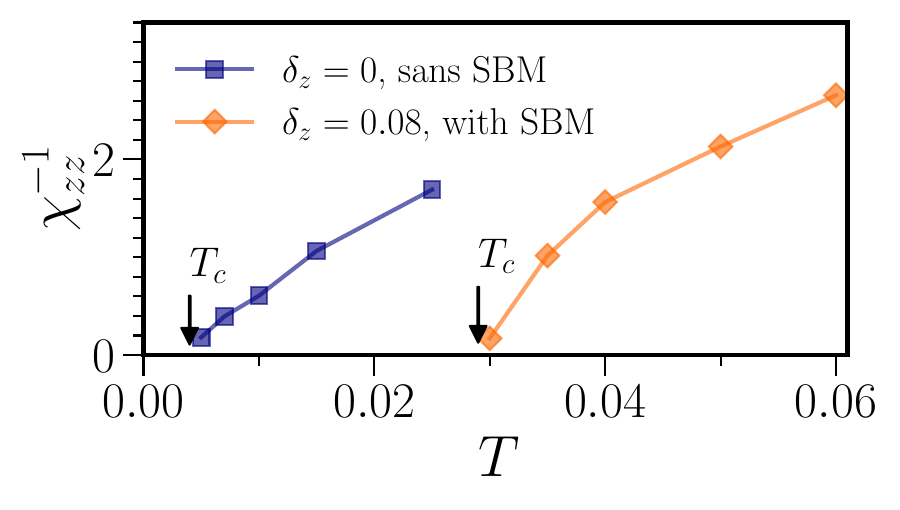}
  \end{center}
  \caption{
    The inverse pairing susceptibility $1/\chi_{zz}$ as a function of temperature $T$ at two hole dopings. Here $\chi_{zz}$ is the inter-layer-intra-orbital  component of the $s_{\pm}$ pairing  susceptibility  within the $d_{z^2}$ orbitals.  Used parameters are $ V=0.5t, t_{\perp} = -0.8t, n_x = 0.6$. For the interaction parameters, $J_H =0, U_{xx}^{\uparrow  \downarrow} = 0, U_{xz}^{\sigma  \sigma^{\prime}} =0, $  $1/\chi_{zz} \rightarrow 0 ,U_{zz}^{\uparrow  \downarrow}=7t$. Note that $\chi_{zz}^{-1} \rightarrow 0 $ at small $T$ indicates the occurrence of  superconducting instability (black arrows).
    }
\label{fig:sbm}    
\end{figure}

In the main text, we have shown that when the  $\sigma-$ bonding band ($\gamma-$ band) metallization (SBM) is absent ($\delta_z=0$),  the superconducting $T_c$ of SC II pairing will be greatly suppressed. We argue that this is because the enhancement of the low-energy $d_{z^2}$ density of state (DOS), which plays a vital role in SC II pairing, is linked to SBM(see Fig.~\ref{fig:illu} and Fig.~\ref{fig:sc} in the main text) in $\mathrm{La_3Ni_2O_7}$ systems.
However, it is worthy noting that, even without SBM, or namely when the $\gamma -$ band of $d_{z^2}$  orbital is under Fermi level, the  $d_{z^2}$ DOS at Fermi level is  non-zero (see Fig.~\ref{fig:illu}), due to the finite hybridization between $d_{x^2-y^2} -$ $d_{z^2}$ orbitals. One natural question then is, can this small but finite  low-energy $d_{z^2}$  DOS in principle give rise to a non-zero superconducting $T_c$ ? The $T_c$ in the absence of SBM could be too small to detect in numerical calculations.  In the main text, the lowest temperature we have accessed is $T=0.008t$ at $\delta_z=0, J_H = 0.15U$, where no signature of SC is found (see Fig.~\ref{fig:sc}). Here we simplify the  $H_U$ part of 2d BLTO Kanamori Hubbard model (Eq. S1) to tackle this issue, namely, neglecting the Hund's coupling $J_H$ and Hubbard repulsions between   $d_{x^2-y^2} -$ $d_{z^2}$ orbitals $U_{xz}^{\sigma  \sigma^{\prime}}$, and repulsions between $d_{x^2-y^2}$ electrons $U_{xx}^{\uparrow  \downarrow}$, keeping only the Hubbard repulsion between $d_{z^2}$ electrons  $U_{zz}^{\uparrow  \downarrow}=7t$. Under such simplification, we can access much lower temperature regime of $T \gtrsim 0.004t$ while the 
SC II pairing is maintained~\cite{zheng2025s}.
Fig.~\ref{fig:sbm} displays  inverse pairing susceptibility $1/\chi_{zz}$ as a function of temperature $T$ for two hole-dopings, $\delta_z = 0$ (without SBM), and $\delta_z = 0.08$ (with SBM). Extrapolating $1 / \chi_{zz}$ to zero, $1 / \chi_{zz} \rightarrow 0$, we identify two $T_c$ for the two 
cases, $T_c \approx 0.004t$  and $T_c \approx 0.03t$ respectively. From this result, we learn that SC II pairing may exist at extremely low temperatures in the absence of SBM. Its transition $T_c$ is, however, greatly suppressed comparing to the case with SBM. It is also much lower than the $T_c$ of SC I pairing at  $\delta_z = 0$ under an enhanced $J_H$ ($T_c \approx 0.017t$ at $J_H = 0.2U$).

\section*{F\quad Evolution of the magnetic correlations}
The unconventional superconductivities in $\lno$ are believed to ultimately  connects to the magnetic correlations~\cite{wu24_charge} in the system. Here we present result on static magnetic correlator  $\langle S  \cdot S \rangle$  as a function of  $\dz$ hole-doping level $\delta_z$. 
\begin{figure}
  \begin{center}
    \includegraphics[width=1\columnwidth]{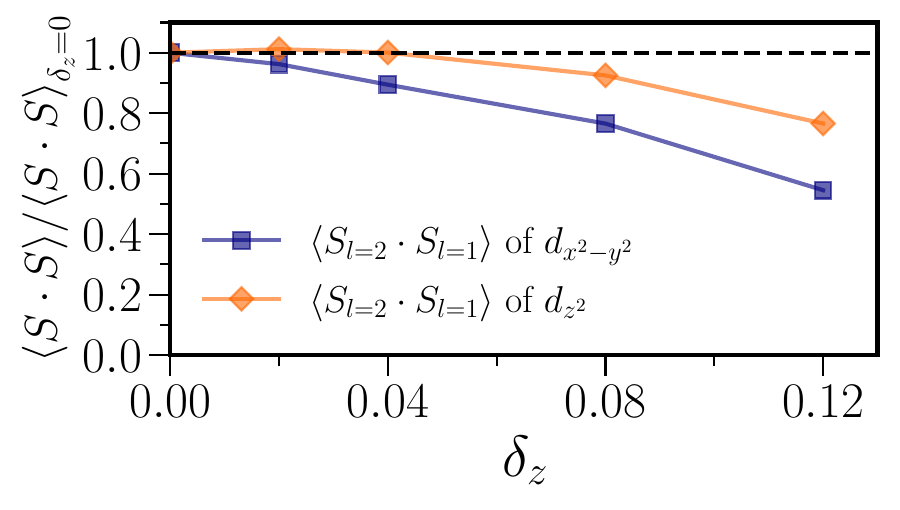}
  \end{center}
  \caption{
    The interlayer intra-orbital static magnetic correlator $\langle S_{\alpha,l=2} \cdot S_{\alpha,l=1}\rangle$  as a function of $\delta_z$ for $\alpha = \dx$  and $\alpha = \dz$ orbitals. Here $T=0.02t$, other parameters are  $ V=0.5t, t_{\perp} = -0.8t, n_x = 0.6, n_z = 1- \delta_z, J_H = 0.2U$.
    }
\label{fig:ss}    
\end{figure}

Fig.~\ref{fig:ss} displays the magnetic correlator $\langle S_{l=2} \cdot S_{l=1}\rangle$ between  interlayer $\dx$ orbitals (Squares) and interlayer $\dz$ orbitals (Diamonds) calculated within the $1\times 4-$  orbital impurity cluster of CDMFT (thus this quantity is local in real-space). To better depict the tendency with $\delta_z$, the shown results are renormalized by their corresponding $\delta_z=0$ values, which are $ \approx -0.038$ and  $ \approx -0.448$ respectively for interlayer $\dx$ and interlayer $\dz$ orbitals at $T=0.02t$. From this plot we see that as $\delta_z$ increases, both magnetic correlators decrease. The $\dx$ component is, somehow, seen being suppressed at a faster pace than the $\dz$ component, despite  $\delta_z$ is the hole-doping in $\dz$ orbital. We find that the  suppression of superconducting $T_c$ by increasing  $\delta_z$ is not fully synchronized with the reduction of  $\langle S  \cdot S \rangle$ with $\delta_z$. For example, SC I can be fully suppressed when $\delta_z \gtrsim 0.06$ at $J_H = 0.2U$ (see Fig.~\ref{fig:sc} and Fig.~\ref{fig:pd} in the main text), but here  $\langle S  \cdot S \rangle$ remains about $80 \%$ of its $\delta_z = 0$ value at $\delta_z = 0.08$. Similar phenomenon has also been observed in the study of cuprate superconductivity~\cite{wu25interplay}. This may because the static magnetic correlator shown in Fig.~\ref{fig:ss} includes much high-energy contribution. The magnetic correlations driving SC, however, are usually of low/intermediate energy scales~\cite{wakimoto07,gull22}. 



\section*{G\quad Hubbard $U$ effects on SC II at small Hund's coupling $J_H$}
In the main text we have shown that the SC II pairing at $\delta_z = 0.08$ displays a $T_c$ decreasing with $J_H$ when $J_H < 0.2U = 1.4t$. We argue that this is due to the reduced correlation strength of $\dz$ orbitals, but not to the intrinsic effect of $J_H$.
(by "intrinsic effect", we mean the transferring of super-exchange interactions from inter-layer $\dz$ orbitals to inter-layer $\dx$ orbitals by $J_H$, an effect driving the SC I superconductivity.)
Indeed, as shown below in Fig.~\ref{fig:utc}, the superconducting $T_c$ can be enhanced, if we increase  Hubbard $U$ while a small $J_H = 0.7t$ is fixed. At $U =11t, J_H = 0.7t$, $T_c$ can be restored to the value of the case with a large Hund's coupling at $U =7t, J_H = 1.4t$, where $T_c \approx 0.02t$.
\begin{figure}
  \begin{center}
    \includegraphics[width=1\columnwidth]{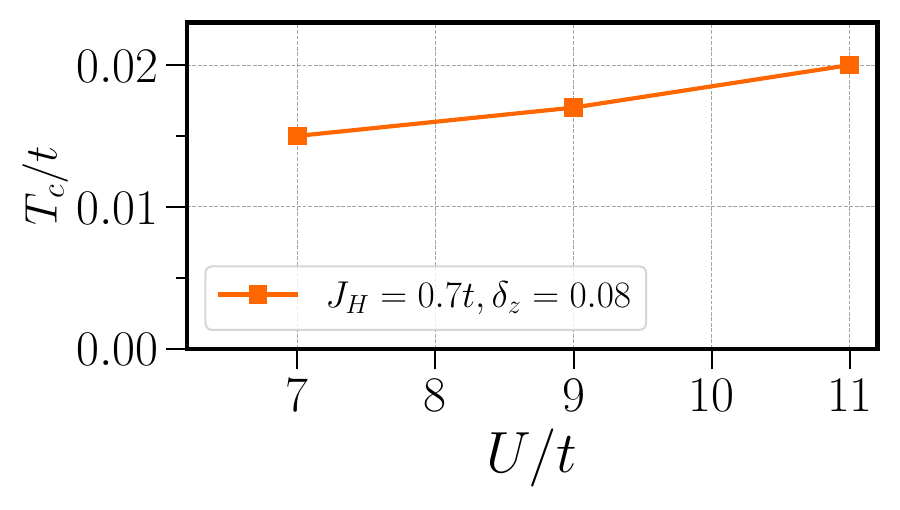}
  \end{center}
  \caption{
      Superconducting $T_c$ as a function of Hubbard $U$ at $J_H = 0.7t$.
    }
\label{fig:utc}    
\end{figure}
\bibliographystyle{apsrev4-2}
\bibliography{ni7}
\end{document}